# A Software Tool for Evaluating Unmanned Autonomous Systems

**Abdollah Homaifar, Ph.D.**

**Ali Karimoddini, Ph.D.**

North Carolina Agricultural and Technical State University, Greensboro, NC

**Mike Heiges, Ph.D.**

Georgia Tech. Research Institute, Atlanta, GA

**Mubbashar A. Khan, Ph.D.**

**Berat A. Erol, Ph.D.**

**Shabnam Nazmi**

North Carolina Agricultural and Technical State University, Greensboro, NC

**Abstract**

*The North Carolina Agriculture and Technical State University (NC A&T) in collaboration with Georgia Tech Research Institute (GTRI) has developed methodologies for creating simulation-based technology tools that are capable of inferring the perceptions and behavioral states of autonomous systems. These methodologies have the potential to provide the Test and Evaluation (T&E) community at the Department of Defense (DoD) with a greater insight into the internal processes of these systems. The methodologies use only external observations and do not require complete knowledge of the internal processing of and/or any modifications to the system under test. This paper presents an example of one such simulation-based technology tool, named as the Data-Driven Intelligent Prediction Tool (DIPT). DIPT was developed for testing a multi-platform Unmanned Aerial Vehicle (UAV) system capable of conducting collaborative search missions. DIPT's Graphical User Interface (GUI) enables the testers to view the aircraft's current operating state, predicts its current target-detection status, and provides reasoning for exhibiting a particular behavior along with an explanation of assigning a particular task to it.*

## Introduction

Advances in sensors and processing technologies have been driving the development of the Unmanned Autonomous Systems (UASs) towards increasing levels of autonomy [Homaifar, et al., 2019]. Rudimentary scripted behaviors, such as GPS waypoint following will be superseded by more advanced capabilities such as search and detect, route de-confliction, collision avoidance, collaborative task assignment, and attack. A significant advancement in the UAS autonomy is also occurring in sensor-based behaviors for navigation, obstacle avoidance, object recognition, path planning, target detection and control. The use of vision-based technologies is also increasing due to the need for UAS to operate in the GPS-denied and communications-degraded environments. As a result, testers need to predict behaviors and evaluate the performance of increasingly intelligent systems, especially those that employ perception-based behaviors.

The dynamic, non-deterministic behaviors of intelligent autonomous systems present the testers with a significant challenge. Testers must predict behaviors that may emerge as the result of the external stimuli or changes in the environment. The key to prediction is thorough testing and evaluation of the complete autonomy process from sensor signal processing through perception, decision-making, action, and response in an immersive simulation environment.

Testing with a high-fidelity simulator can significantly aid in the evaluation of UAS behaviors. Simulations enable a wide variety of mission-scenarios to be investigated safely, and they inherently provide truth data for the comparisons with the UAS's perceptions. The simulation tools have matured to the point that they can readily provide an immersive environment with realistic sensor stimuli for triggering the behaviors of intelligent autonomous systems. In addition to current simulation capabilities, the testers need tools that offer insight into the UAS's decision-making process. If a UAS exhibits an unexpected response, the testers should be able to find the answers to the questions, such as, "Why did it do that?" or "What mode is it operating in now?" The testers need the ability to peer into the autonomy process to identify the divergence in the UAS's decision-making from expected behaviors. Further, the UAS's behavioral modeling enables the testers to predict such occurrences in different scenarios.

NC A&T has addressed these needs through the development of methodologies and the creation of technologies that can monitor the UAS's behaviors, infer its internal states of reasoning, and predict system performance across a wide range of scenarios. The resulting technologies of DIPT use only external observations and do not require any modifications to the System Under Test (SUT). In addition, DIPT is only loosely coupled to the simulation environment and, as a result, is largely simulator agnostic.

**Problem Statement**

The T&E community at the DoD requires insight into a UAS's internal states of reasoning, so the testers can then develop an understanding of the system's behavioral states in response to the particular situation(s). This understanding also helps the tester confidently predict the UAS's behavior in different scenarios. The T&E community might require the system developer to provide access to the system's internal processes for their monitoring during the testing process of the UASs that are being developed under the DoD's funding. However, the commercial sector is quickly dominating the development of advanced autonomy due to the availability of a far larger (and more competitive) market than the military market for it. The result is autonomy with greater capabilities and lower costs. A current example of this trend is the development of self-driving cars. Their development was initially spurred by the DARPA Grand Challenge and Urban Grand Challenge but has now transitioned to companies such as Google and Uber, which are further developing it for the commercial market. While the military would like to take advantage of the lower costs and greater flexibility of these commercial products, the drawback is these commercial products are not designed with the *tester's* needs in mind. In the case of self-driving cars, the military may wish to use commercially developed driverless car technology for automated convoys, but they will eventually be faced with the challenge of evaluating a highly autonomous system that is based on proprietary software with little or no access to its internal processes. Without the ability to gain insight into the UAS's decision-making processes, the testers will be unable to predict its behavior in a wide variety of scenarios. This limits the testers' ability to develop and execute safe and appropriate test plans.

The other challenge that testers face with advanced UASs is the proper stimulation of all aspects of the system that may affect its decision-making processes. This includes not only the sensors used for monitoring the UAS's states (such as speed, heading, and location, etc.), but also those that are used for monitoring its environment (such as video cameras, radar, and LIDAR, etc.) and communication inputs from the operators and/or other autonomous vehicles. Because all of the processing steps from the inputs to the sensors and communications to the actuator commands can affect the system's behavior, they must all be stimulated to fully exercise the system's autonomy. Thus, simulation-based testing requires an immersive simulation environment to stimulate the UAS's autonomy over all input space.

**Technical Approach**

NC A&T's approach to the development of DIPT's Inference Engines (IEs) was based on the investigation of two of key technologies used for inferring behaviors and modeling the rules, namely, the Learning Classifier Systems (LCSs) [Butz, 2015] and the Type-2 Fuzzy Logic Systems (FLSs) [Mendel, 1995; Hailemichae, et al., 2018]. An LCS is a machine learning paradigm based on reinforcement learning and Genetic Algorithms (GAs) in which a system learns to perform a certain task by interacting with a partially known environment via the guidance of a reward signal that indicates the quality of its action [Workineh and Homaifar, 2011]. It is an expert system that utilizes a knowledge base of syntactically simple production rules that can be manipulated by the GA [Shafi and Hussein, 2017]. The use of a rule-based system allows an LCS to conveniently represent the complex control strategies [Karlsen and Sotiris, 2018]. The LCS has been used to develop an IE that deduces the behaviors of the UAS as its current operating state by observing its external actions [Workineh and Homaifar, 2012]. Some other related research about the LCS can be found in listed references [Urbanowicz and Jason, 2009; Urbanowicz and Jason, 2014; and. Chang and Sung, 2017].

The other technology, the Type-2 Fuzzy Logic, was used to infer the UAS's perception of its environment through a model of the input/output relationships [Qilian and Mendel, 2000]. In the fuzzy systems, an expert's heuristic knowledge is translated into the fuzzy IF-THEN statements, which, along with an appropriate IE, are used for system identification [Enyinna, et al., 2015]. In comparison to the ordinary fuzzy systems, the fuzzy Type-2 provides a more plausible way to reduce the effect of uncertainty in the rule-based FLSs [Hailemichael, et al., 2019]. The uncertainties inherent in the antecedents and consequences of the rules, the expert knowledge-base, the sensor measurements, and the system responses can be modeled with additional degrees of freedom using the degree of uncertainty of the Fuzzy Membership Functions (MFs) [Wu, 2013]. Recent literature about T2 fuzzy logic and its applications is available in works by Almaraashi, et al., 2016; Melin, 2018; Antão, et al., 2018; Bennaoui and Slami, 2017; and Hailemichael. et al., 2018 referenced at the end of this article.

To focus the research effort on the development of the IEs using these two methodologies, a representative use case of an autonomous system with advanced capabilities was selected: GTRI's Collaborative Unmanned Systems Technology Demonstrator (CUSTD) [Pippin, et al., 2010]. CUSTD includes multiple UAVs that are capable of collaboratively searching an area for a ground target using onboard vision processing. Once a team member locates a potential target based on the collected information, either it has the required equipment to verify the target or it requests the other team members for the verification of the existence of the target.

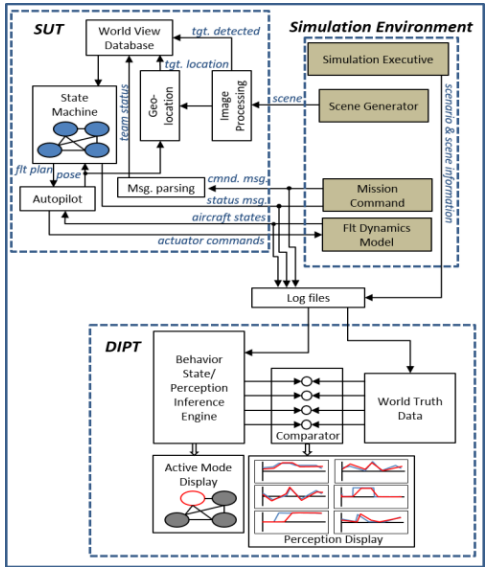

*Figure 1: DIPT-SUT-Simulation Architecture*

When developing the methodologies to infer a UAS's perception and behavior-process, the challenge is to validate the output of the resulting IE for the UASs for which one does *not* have the ability to modify the code to output the behavior state and perception information. By using GTRI's CUSTD system, the internal logic that controls the UAS's behaviors was made accessible and used for the validation of the models derived by DIPT. This ensured that the methodology developed for DIPT could be applied to the other UASs performing an Intelligence, Surveillance, and Reconnaissance (ISR) mission.

The way in which DIPT is used with a UAS (the SUT) and the simulation environment is illustrated in Figure 1. As indicated in the figure, the data that is external to the UAS's autonomy processes is written into the log files. The data that is currently available in the autopilot's telemetry stream includes the aircraft states such as the vehicle's location, heading, heading rate, and airspeed. These can be observed for clues about the UAS's active behavior mode. Information describing the simulation scenario is also written into the log files. This information includes data on the characteristics of the camera, such as its field of view, the coordinates of the search area, the location of the target, visibility conditions, and light level, etc. Some of this information is used by IEs while some of it is used as the truth data in order to compare it with the UAS's perception. For example, the Comparator function compares where the UAS thinks a target is located versus its actual location as modeled in the simulator. The evaluation tool includes a GUI for the testers to display the comparisons between the UAS's perceptions and truth data in real time. The interface also shows the inferred active behavioral state and the status of the triggers that control the transitions to other states.

**Results**

The first step in the development of the LCS for inferring the UAS's internal states and state transition triggers was to create a representative state machine of the GTRI CUSTD system using an educated guess. This approximated model was based on the information gathered from published papers about the CUSTD system and on the inputs from subject-matter experts. The assumed state machine is presented in Figure 2 while the state transition triggers are presented in Table 1. A hierarchical LCS model was built in which the task of the learning system was to categorize the states into groups that can be treated in a similar way. The hierarchical model allows for a default hierarchy that refers to the ruleset with a default rule for the default class along with others as the exception rules. In the hierarchical ruleset, the accurate and more specific rules respond to a subset of the situations covered by more general but less accurate default rules. Unlike the homomorphic LCS models that require a vast number of rules to model the realistic environment, the default hierarchy allows the building of a more compact ruleset with reasonably fair accuracy.

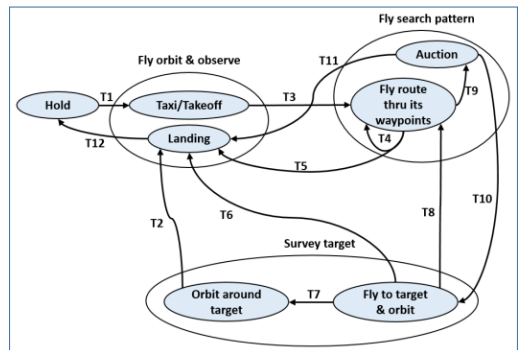

*Figure 2: Assumed CUSTD State Machine*

The system's performance improves with the increasing number of more exception rules in the hierarchy. The set of correct transition rules were published by the system developer C. Pippin, et al., 2010 and used for training the LCS block as a known ruleset. For example, the learning behavior of state transitions for the UAV with states represented by Figure 3, typical examples of rules are as follows:

*Table 1: State Transition Triggers*

| Trigger | Description |
|---|---|
| T1 | Start mission |
| T2 | Mission aborted |
| T3 | The first waypoint reached |
| T4 | Repeat (Target not detected & end of search pass number not reached) |
| T5 | Target not detected & reached the end of search pass number |
| T6 | Target detected with low confidence & reached the end of search pass number |
| T7 | Target detected with high confidence |
| T8 | Target detected with low confidence & end of search pass number not reached |
| T9 | Target detected |
| T10 | Primary UAV wins the auction |
| T11 | Primary UAV losses the auction |
| T12 | Landing confirmed |

- **Rule 1:** If the current state of the UAS is at "Fly-Search-Pattern" and a target marker is detected, then switch or move to "Survey Target" state.
- **Rule 2:** If the current state is at "Survey-Target "and a "Return" signal is received, then move to "Fly-Orbit-and-Observe" state.
- **Rule 3:** If the current state is at "Fly-Search-Pattern" and "Help is Requested" from other UASs in the team at location (x, y), then switch to "Survey-Target" state.

The developed LCS model was trained (using the variables that define the aircraft's operation) to recognize the patterns in the aircraft's operation in each state of the state machine. The LCS algorithms provide multi-threading capability and facilitate the simultaneous and parallel processing of the jobs.

Multiple approaches have been implemented and integrated with the LCS training algorithm. Compact Rule-set Algorithm 2 (CRA2) and Parameter-Driven Rule Compaction (PDRC) have the largest rule-reduction and the least performance deterioration. A two-layer MapReduce structure has been employed in case in which a current algorithm fails to handle a large number of scenarios. The LCS was used for the prediction of correct state transitions for a previously unknown input scenario.

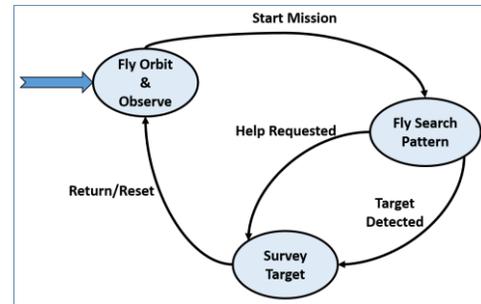

*Figure 3: UAV Behavior-Based Autonomy*

To improve the prediction performance of the developed LCS model, new features were introduced by combining the existing ones, mathematically. Two such features are calculated by a nonlinear combination of the linear velocities of the UAV. The simulations confirm the positive effect of considering such features. Table 2 shows the inference of the transition triggers by the developed LCS model. There is only one transition which is designed to take place with multiple possible triggers. A binary classifier proved to be the proper model for the discrimination between 'search timeout reached' and 'search complete' triggers.

Training an LCS model for different UAV groups required fine-tuning of a few parameters, such as the number of training iterations, training instances, and overall population size. In order to meet a desired accuracy for a UAV group called Osprey Beta III, the optimal training size and the training iterations were found to be 5000 and 120000, respectively. The model behaved as expected and reached the desired accuracy rates when the training set size (the number of training instances) is around 9000. Keeping the size of the training set fixed at 9,000, the highest accuracy rate was achieved with a population size of 16,000. Keeping the training size fixed at 9,000 and the population size at 16,000, it was found that 15,000 iterations were required to achieve the optimal accuracy.

*Table 2: State Transition Triggers for LCS Model*

| From | To | Trigger |
|---|---|---|
| Hold | Fly Orbit and Observe | Go for launch |
| Fly Orbit and Observe | Fly Search Pattern | First search waypoint reached |
| | Hold | Landing complete |
| Fly Search Pattern | Survey Target | Potential target found, and auction won |
| | Fly Orbit and Observe | Search timeout reached |
| | | Search complete |
| | | Battery low |
| | | Abort mission |
| | Fly Search Pattern | Potential target found, and auction lost |
| Survey Target | Fly Orbit and Observe | Survey complete |
| | | Battery low |
| | | Abort mission |
| | Fly Search Pattern | Potential target lost |

The FLS developed for the inference of the UAS's perception of its environment has been based on a Type-2 FLS [Liang and Mendel, 2000]. It consists of the data preprocessing, an FLS engine, and an FLS tester's display. The data preprocessing block takes input from the CUSTD simulation and performs operations such as deriving parameters from a set of basic parameters, merging parameters of different sampling rates, and normalization. The parameters are categorized into four classes, namely the UAV parameters, the environmental parameters, the imagery characteristics, and the target information.

The FLS-based perception engine consists of a fuzzifier, an inference drawing block, and output processing and the comparator blocks that are developed based on the MFs and rules, as shown in Figure 4. The first step in FLS development is determining MFs. Based on the CUSTD simulation data and experiment set up, appropriate MFs for each input were chosen. Then rules representing the UAV's performance for all possible permutations of antecedents were collected from experts and extracted from published image processing papers. The FLS engine was tested with inputs from the newly generated scenarios. The output of the inference processes is displayed on a Graphical User Interface (GUI), as shown in Figure 5. The GUI displays the currently active mode for the UAS as well as a running comparison of its perception of the world state with the actual world state as provided by the simulation environment. A preliminary FLS-report was generated for the analysis of the UAS flight, UAS perception, and FLS perception and the rules to give the test engineers an insight as to why a UAV fails to detect the target.

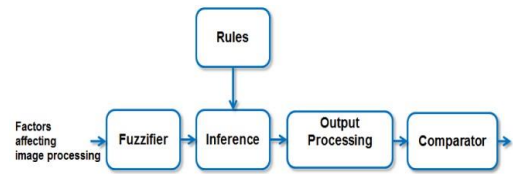

*Figure 4: The FLS implementation*

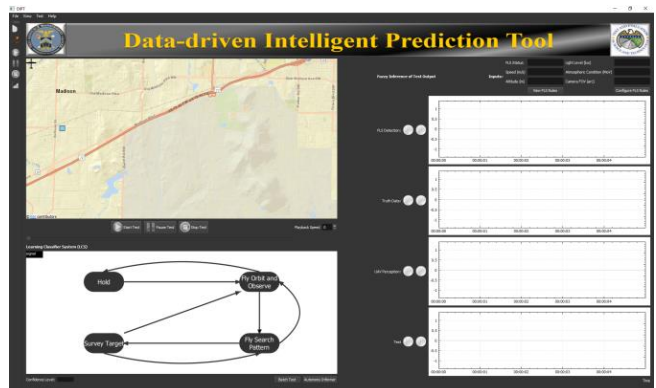

*Figure 5: The GUI for DIPT*

For example, once the CUSTD UAV detects a target marker, it determines the marker's geolocation coordinates. These coordinates are compared to those of the marker's actual coordinates (as specified in the simulation) and displayed on the GUI to help the tester in the evaluation of the accuracy of the UAS's geolocation algorithm.

**Conclusions**

With DIPT, NC A&T has demonstrated the use of the two methodologies, LCS and FLS to infer the operating state and perceptions of a UAS using only the externally available information and observations. DIPT allows testers to deduce such occurrences and provides the associated reasoning in different scenarios. A step-by-step process for initially establishing and then refining an LCS rule-base in an efficient automated manner was developed. The initial results from a use-case based on the GTRI CUSTD UAS are very promising. For modeling a UAS's perception-behavior, NC A&T showed how a Type-2 FLS accommodates the variations in the models proposed by multiple subject matter experts. This technique was used successfully to create a rule-base for modeling the perception performance of the CUSTD UAS. A set of CUSTD simulation scenarios were tested using the FLS and the FLS was found to be capable of classifying the UAS perception performance very well. The DIPT-GUI presents a situational display to allow the tester to view the scenario test results as they unfold. The simulation data is stored in log files that the tester can choose to pause and replay the portions of the simulation. The situational display shows the location and track of the UAV as it conducts a search mission.

**Acknowledgements**

*The authors acknowledge and thank* the Test Resource Management Center (TRMC) and Test Evaluation/Science & Technology (T&E/S&T) Program and/or the US Army Contracting Command Orlando (ACC-ORL-OPB) for funding the *Data-driven Intelligent Prediction Tool (DIPT) project under contract W900KK-17-C-0002*. Any opinions, findings, and conclusions or recommendations expressed in this material are those of the author(s) and do not necessarily reflect the views of the Test Resource Management Center (TRMC) and Test Evaluation/Science & Technology (T&E/S&T) Program and/or the US Army Contracting Command-Orlando (ACC-ORL-OPB). Also, the first and the second author would like to acknowledge the partial support by NASA University Leadership Initiative (ULI) under agreement number 2 CFR 200.514.

---

*ABDOLLAH HOMAIFAR, Ph.D., is the NASA Langley Distinguished Chair Professor and the Duke Energy Eminent Professor at North Carolina A&T State University (NCA&TSU). He is the director of the Autonomous Control and Information Technology Institute and the Testing, Evaluation, and Control of*

*Heterogeneous Large-scale Systems of Autonomous Vehicles centers at NCA&TSU. His research expertise includes unmanned autonomous systems, intelligent robotics, machine learning, optimal control, soft computing, and modeling. Among his current sponsored research are grants from the US Department of Defense, Air Force Research Laboratory, Lockheed Martin Corporation, and the Office of Under Secretary of Defense Test Resource Management Center.*

*ALI KARIMODDINI, Ph.D., is the director of NC-CAV Center of Excellence on Advanced Transportation, the director of the Autonomous Cooperative Control of Emergent System of Systems (ACCESS) laboratory, and the Deputy Director of the TECHLAV DoD Center of Excellence in Autonomy at N.C. A&T State University. His research interests include Control and Robotics, Flight Control Systems, Human-machine Interactions, Cyber-physical Systems, and Multi-agent Systems. He has received more than $25M prior support from federal funding agencies and industrial partners to research development of autonomous vehicles and applications.*

*MIKE HEIGES, Ph.D., is a Principal Research Engineer at the Georgia Tech Research Institute where he serves as the Associate Division Chief for the Robotics and Autonomous Systems Division. His background is in aircraft flight dynamics and automatic control and he manages several of GTRI's swarming UAV programs. Dr. Heiges earned his Ph.D. from the Georgia Tech School of Aerospace Engineering in 1989. He is an Associate Fellow of the American Institute of Aeronautics and Astronautics (AIAA) and a member of the Association for Unmanned Vehicle Systems International (AUVSI).*

*MUBBASHAR A. KHAN, Ph.D., is a Project Manager of the North Carolina Transportation Center of Excellence on Connect and Autonomous Vehicle Technology (NC-CAV) at North Carolina A&T State University. In the past, Dr. Khan worked as a Post-Doctoral Scholar at North Carolina A&T State University and as an Assistant Professor at Mirpur University of Science and Technology, Pakistan. Dr. Khan holds a Ph.D. degree in Engineering from the University of Toledo (2018). Dr. Khan's research interests include Fuzzy logic and Type-2 Fuzzy Systems, autonomous navigation, Human-Machine Interactions, Cognitive Radios, QoS and secondary radio spectrum, and Machine Learning Algorithms.*

*BERAT A. EROL, Ph.D., is a Post-doctoral research associate at the ACIT Institute. Before joining ACIT, he was with the Autonomous Control Engineering Labs at the University of Texas at San Antonio (UTSA), where he earned his Ph.D. degree in 2018. His doctoral work focused on autonomous systems, human-robot interactions, and manned-unmanned teaming. Dr. Erol's dissertation was supported by several research grants and contracts that he contributed to, including the US DoD, Bank of America, 80/20 foundation, UTSA Lutcher Brown Endowed Chair, and UTSA Open Cloud Institute. He is member of IEEE, AIAA and IEEE HKN honor society.*

*SHABNAM NAZMI, is a Ph.D. candidate at the ACIT institute at the Department of Electrical and Computer Engineering, North Carolina A&T State University. She received her B.S. degree in Electrical Engineering from K.N.Toosi University of Technology and her M.S. degree in Electrical Engineering from Sharif University of Technology in 2009 and 2012, respectively. Her research interests include test and evaluation of autonomous vehicles, multi-label classification and its application to test and evaluation of autonomous vehicles, genetic-based machine learning, and learning from streaming data.*